\documentclass[10pt,a4paper]{article}

\newcommand{\iprime}{{i^\prime}}
\newcommand{\jprime}{{j^\prime}}
\newcommand{\kprime}{{k^\prime}}
\newcommand{\lprime}{{l^\prime}}

\newcommand{\rhoprime}{{\rho^\prime}}

\begin{document}

\title{Space and Time models\\
{\normalsize Mastering General Covariance}}
\author{Ll. Bel\thanks{e-mail:  wtpbedil@lg.ehu.es}}

\maketitle

\begin{abstract}

We derive line-element's templates of space-time models with Space models complying  with Helmholtz's Free mobility postulate, and discuss some of the Time models compatible with them.

\end{abstract}

\section*{Introduction}

General relativity owns much to the continuous efforts made by mathematicians to understand the logical coherence of Euclide's postulates that led them to discover new geometries. Helmholtz, a physician and physicist, introduced in this quest the empiricist idea that the geometry of space should be such that rigid solids could be displaced and rotated without modifying their geometrical intrinsic properties. This ``Free mobility postulate" led him to the conclusion that a 3-dimensional space geometry should be a Riemannian space with constant curvature\,\footnote{ As a short introduction to the subject I suggest reading the web pages {\it Reichenbach on Helmholtz}, {\it Riemann and Helmholtz}, and {\it Universalgenie Helmholtz}.
Ref.\cite{Russell} is a full book on the subject.}
.

General relativity is a far-reaching formal theory of space-time as well as a successful theory of gravitation where by geometry is meant essentially 4-dimensional space-time geometry, and this raised, from the very beginning, fundamental difficulties to deal with rigid bodies. As a consequence Helmholtz's ideas are considered now by most as completely obsolete.

This paper is a proposal to re-consider Helmholtz's empiricist point of view in the frame-work of General relativity implementing the Free mobility postulate to define the  Space models and the Time models that are compatible with them.

\section{Frames of reference}

Let:

\begin{equation}
\label{X.1}
ds^2=g_{\alpha\beta}(x^\rho)dx^\alpha dx^\beta
\end{equation}
be the line-element of a space-time model referred to an arbitrary system of coordinates $(t,\ x^i;\ i,j,k,\dots=1,2,3)$. And let $u^\alpha$ be a unit time-like vector field. The congruence $\cal C$ that it defines will be interpreted as a frame of reference if a definition of space distance can be found such that the distance between world-lines remains constant.

Congruences able to to describe a frame of reference must satisfy accepted geometrical postulates while congruences describing the evolution of a physical system must satisfy the laws governing it. This paper is focused in a proposition to lay down the postulates that a frame of reference has to satisfy.

A system of coordinates is said to be adapted to the congruence $\cal C$ defined by $u^\alpha(x^\rho)$ if:

\begin{equation}
\label{X.2}
u^i=0
\end{equation}
the coordinate transformations that leave invariant these conditions are ($t=x^0$):

\begin{equation}
\label{X.3}
t=t(t^\prime,x^\jprime),\quad x^i=x^i(x^\jprime)
\end{equation}
and can be therefore be decomposed into a pure adapted time transformation;

\begin{equation}
\label{X.4}
t=t(t^\prime,x^j), \quad x^i=x^i
\end{equation}
and a pure adapted space transformation:

\begin{equation}
\label{X.5}
t=t\quad x^i=x^i(x^\jprime)
\end{equation}

Intrinsic properties of time-like congruences are those that are invariant under adapted coordinate transformations.

Under adapted coordinate transformations we have in particular the following relationship:

\begin{equation}
\label{F.1}
\Gamma^\iprime_{\alpha^\prime\beta^\prime}(x^\rhoprime)g^{\alpha^\prime\beta^\prime}=
A^\iprime_j\Gamma^j _{\mu\nu}g^{\mu\nu}(x^\rho)-g^{jk}\partial_jA^\iprime_k
\end{equation}
where the $\Gamma$´s are the Christoffel symbols and:

\begin{equation}
\label{F.2}
A^\iprime_j=\frac{\partial x^\iprime}{\partial x^j}, \quad
A^i_\jprime=\frac{\partial x^i}{\partial x^\jprime}
\end{equation}
Let us assume that we know how to associate to the congruence being considered a 3-dimensional time independent connection $\tilde\Gamma$ that transforms concomitantly to $\Gamma$ under adapted coordinate transformations:

\begin{equation}
\label{F.3}
\tilde\Gamma^\iprime_{\jprime\kprime}(x^\lprime)=A^\iprime_s\tilde\Gamma^s_{mn}(x^l)A^m_\jprime A^n_\kprime+
A^\iprime_s\partial_{j^\prime}A^s_{k^\prime}
\end{equation}
We can then construct the following subordinate quantities:

\begin{equation}
\label{F.4}
L^i\equiv \Gamma^i_{\mu\nu}g^{\mu\nu}-\tilde\Gamma^i_{jk} g^{jk}
\end{equation}
that transform as the components of a vector of the quotient manifold defined by the congruence $\cal C$:

\begin{equation}
\label{F.5}
L^\iprime=A^\iprime_jL^j
\end{equation}

Using a system of coordinates adapted to $u^\alpha$ the line-element (\ref{X.1}) becomes:

\begin{equation}
\label{0.1}
ds^2=-A(t,x^i)^2(-dt+f_j(t,x^i)dx^j)^2 \\ [1ex]+A(t,x^i)^{-2}d\bar s^2
\end{equation}
where:

\begin{equation}
\label{X.6}
A=\sqrt{-g_{00}},\quad f_j=A^{-2}g_{0j}
\end{equation}
and:

\begin{equation}
\label{0.2}
d\bar s^2=\bar g_{jk}dx^jdx^k,\quad \bar g_{jk}=A^2(g_{jk}+ A^2f_jf_k)
\end{equation}

Let us consider an adapted time-transformation (\ref{X.4}). Under such a time transformation we have:

\begin{equation}
\label{0.5}
A^\prime(t^\prime,x^i)=A\frac{\partial t}{\partial t^\prime}
\end{equation}
Similarly we have:

\begin{equation}
\label{0.6}
\bar g^\prime_{ij}(t^\prime,x^k)=\bar g_{ij}\left(\frac{\partial t}{\partial t^\prime}\right)^2, \quad
\bar g^{\prime ij}(t^\prime,x^k)=\bar g^{ij}\left(\frac{\partial t}{\partial t^\prime}\right)^{-2}
\end{equation}
and also:

\begin{equation}
\label{0.7}
f^\prime_i(t^\prime,x^j)=\left(\frac{\partial t}{\partial t^\prime}\right)^{-1}\left(f_i-\frac{\partial t}{\partial x^i}\right)
\end{equation}
All these quantities are tensors under adapted space coordinates transformations (\ref{X.5}).

Continuing this analysis a short calculation proves that:

\begin{equation}
\label{0.9}
\hat\partial_k \bar g^\prime_{ij}=
\left(\frac{\partial t}{\partial t^\prime}\right)^2\hat\partial_k \bar g_{ij}
+2\bar g_{ij}\frac{\partial t}{\partial t^\prime}X_k
\end{equation}
where:

\begin{equation}
\label{0.10}
\hat\partial_k \bar g_{ij}=\frac{\partial \hat g_{ij}}{\partial x^k}+f_k\frac{\partial \hat g_{ij}}{\partial t}
\end{equation}
and:

\begin{equation}
\label{0.11}
X_k=\frac{\partial^2t}{\partial t^\prime\partial x^k}+
\left(f^\prime_k-\frac{\partial t}{\partial x^k}\right)\frac{\partial}{\partial t^\prime}
\ln\frac{\partial t}{\partial t^\prime}
\end{equation}

But we have also:

\begin{equation}
\label{X.7}
\partial_{t^\prime} f^\prime_i=\partial_t f_i -\left(\frac{\partial t}{\partial t^\prime}\right)^{-1}X_i
\end{equation}
and therefore (\ref{0.9}) can be written:

\begin{equation}
\label{X.8}
\bar\partial_k \bar g^\prime_{ij}=
\left(\frac{\partial t}{\partial t^\prime}\right)^2\bar\partial_k \bar g_{ij}
\end{equation}
where:

\begin{equation}
\label{X.9}
\bar\partial_k \bar g_{ij}=\frac{\partial \bar g_{ij}}{\partial x^k}+f_k\frac{\partial \bar g_{ij}}{\partial t}+2\bar g_{ij}\partial_t f_k
\end{equation}

Let us define the following connection-like symbols:

\begin{equation}
\label{0.12}
\bar\Gamma^i_{jk}=\frac12\bar g^{il}(\bar\partial_j \bar g_{kl}+ \bar\partial_k \bar g_{jl}-\bar\partial_l \bar g_{jk})
\end{equation}
Taking into account (\ref{0.6}) and (\ref{X.9}) we get under an adapted time transformation:

\begin{equation}
\label{0.13}
\bar\Gamma^\iprime_{\jprime\kprime}=\bar\Gamma^i_{jk}
\end{equation}
and therefore they transform as connection symbols under general adapted coordinate transformations (\ref{X.3}).

Let us consider now a 3-dimensional constant curvature Riemannian metric with coefficients independent of the time-variable $t$:

\begin{equation}
\label{0.14.1}
d\tilde s^2=\tilde g_{ij}(x^k)dx^idx^j, \quad  \tilde R_{ij}(x^l)=k\tilde g_{ij}
\end{equation}
We shall say that the line-element (\ref{0.1}) is an implementation of Hemholtz's free mobility postulate with the geometry described by the space metric (\ref{0.14.1}) if the following conditions, which are invariant under any adapted coordinate transformations, are satisfied:

\begin{equation}
\label{0.15}
(\bar\Gamma^i_{jk}-\tilde\Gamma^i_{jk})\bar g^{jk}=0
\end{equation}
where $\tilde\Gamma^i_{jk}$ are the connection symbols of $d\tilde s^2$.

The consideration of space-time models in General relativity implementing Helmholtz's postulate serves two main purposes: i) To allow to consider a frame of reference
with congruence $\cal C$ as a standard of rest by defining the distance between two world-lines
with space-coordinates $x_1^i$ and $x_2^i$ as the integral of $d\tilde s$ above between these two points of space along the geodesic joining them, and ii)
to give a meaning to space-coordinates by saying that they mean for the line-element (\ref{0.1}) the same thing that they mean to (\ref{0.14.1}).


\section {Space models}

Using arbitrary coordinates, let $(a,b,c,\dots=1,2,3)$

\begin{equation}
\label{3.1}
d\tilde s^2=\delta_{ab}\theta^a\theta^b \quad \hbox{with}\quad \theta^a=\theta^a_i(x^k)dx^i
\end{equation}
be a normalized orthogonal decomposition of the line-element (\ref{0.14.1}).
From the definition of the Cartan connection of a Riemannian metric, we have:

\begin{equation}
\label{3.4}
d\theta^a+\tilde\omega^a_b(x^j)\wedge\theta^b=0, \quad
\delta_{ac}\tilde\omega^c_b+\delta_{bc}\tilde\omega^c_a=0
\end{equation}

Writing:

\begin{equation}
\label{3.5}
d\theta^a=\frac12 C^a_{bc}(x^j)\theta^b\wedge\theta^c,
\end{equation}
and using the following notation:

\begin{equation}
\label{3.6}
\tilde\omega^a_b=\tilde\gamma^a_{bc}(x^j)\theta^c,
\end{equation}
a well known procedure leads to the following result:

\begin{equation}
\label{3.7}
\tilde\gamma_{a,bc}=\frac12(C_{a,bc}+C_{b,ca}-C_{c,ab})
\end{equation}
where:

\begin{equation}
\label{3.8}
\tilde\gamma_{a,bc}=\delta_{ad}\gamma^d_{bc}, \quad C_{a,bc}=\delta_{ad}C^d_{bc}
\end{equation}

Let us consider the 3-dimensional metric  (\ref{0.2}) assuming that it can be written as:

\begin{equation}
\label{3.9}
d\bar s^2=\alpha_a^2(t,x^j)\delta_{ab}\theta^a\theta^b
\end{equation}
The corresponding Cartan connection will be now the valued 1-form defined by the two equations:

\begin{equation}
\label{3.10}
d\theta^a+\bar\omega^a_b(t,x^j)\wedge\theta^b=0, \quad
\delta_{ab}\hat d\alpha_a^2=\alpha_c^2\delta_{ac}\bar\omega^c_b+\alpha_c^2\delta_{bc}\bar\omega^c_a
\end{equation}
and a similar calculation that led to (\ref{3.7}) leads now to:

\begin{equation}
\label{3.11}
\bar\gamma_{a,bc}(t,x^j)=\frac12(\bar C_{a,bc}+\bar C_{b,ca}-\bar C_{c,ab})+
\frac12(\partial_b\alpha_a^2\delta_{ac}+\partial_c\alpha_a^2\delta_{ab}
-\partial_a\alpha_b^2\delta_{bc}
\end{equation}
where here:

\begin{equation}
\label{3.12}
\bar\gamma_{a,bc}=\alpha_a^2\delta_{ad}\bar\gamma^d_{bc}, \quad
\bar C_{a,bc}=\alpha_d^2\delta_{ad}C^d_{bc}
\end{equation}
and:

\begin{equation}
\label{3.13}
\hat\partial_a=e^i_a(\partial_i +f_i\partial_t) \quad \hbox{with} \quad  e^i_a\theta^a_j=\delta^i_j
\end{equation}
To implement the free mobility conditions (\ref{0.15}), the following three equations have to be satisfied:

\begin{equation}
\label{3.14}
\partial_a\ln\frac{\alpha_{b}\alpha_{c}}{\alpha_a}+ e^i_a\partial_t f_i=C_{b,ba}(1-\alpha_a^2\alpha_b^{-2})+
C_{c,ca}(1-\alpha_a^2\alpha_c^{-2})
\end{equation}
where $b=a+1$ and $c=b+1$ are defined modulo 3.

It is also useful to introduce the three functions:

\begin{equation}
\label{3.14.1}
h_a(t,x^j)=\frac{\alpha_b\alpha_c}{\alpha_a}, \quad \alpha_a^2=h_bh_c
\end{equation}
in which case the preceding equation becomes:

\begin{equation}
\label{3.14.2}
\partial_a h_a+h_a e^i_a\partial_t f_i=C_{b,ba}(h_a-h_b)+C_{c,ca}(h_a-h_c)
\end{equation}

Below we give a few examples assuming that the constant curvature $k$ is zero, that neither $f_i$ nor the metric (\ref{3.9}) depend on $t$, and that the coordinates diagonalize the Euclidean metric $d\tilde s^2$. The general formulas in this case are:

\begin{equation}
\label{3.14.3}
d\bar s^2=Q_1(u,v,w)^2du^2+Q_2(u,v,w)^2dv^2+Q_3(u,v,w)^2dw^2
\end{equation}
with:

\begin{equation}
\label{3.14.4}
\partial_u h_1=-(h_1-h_2)\partial_u\ln Q_2-(h_1-h_3)\partial_u\ln Q_3
\end{equation}
plus the two cyclic permutations of $u,v,w$ and 1,2,3.

If the coordinates are Cartesian then:

\begin{equation}
\label{3.15}
\theta^a_i=\delta^a_i
\end{equation}
and the line-element (\ref{3.9})becomes:

\begin{equation}
\label{3.16}
d\bar s^2=h_2h_3dx^2+h_3h_1dy^2+h_1h_2dx^3
\end{equation}
with:

\begin{equation}
\label{3.16.1}
\delta_{ia}\partial_ih_a=0
\end{equation}

If the coordinates are cylindrical:

\begin{equation}
\label{3.16.2}
\theta^1=d\rho,\ \ \theta^2=\rho d\varphi,\ \ \theta^3=dz,
\end{equation}
then the line-element (\ref{3.9}) becomes:

\begin{equation}
\label{3.16.3}
d\bar s^2=h_2h_3d\rho^2+h_3h_1\rho^2d\varphi^2+h_1h_2dz^2 \quad
\end{equation}
with:

\begin{equation}
\label{3.16.4}
\partial_\rho h_1=-\rho^{-1}(h_1-h_2)=0,\ \ \partial_\varphi h_2=0, \ \ \partial_z h_3=0
\end{equation}

If spherical coordinates are used:

\begin{equation}
\label{3.16.5}
\theta^1=dr,\ \ \theta^2=rd\theta,\ \ \theta^3=r\sin \theta d\varphi,
\end{equation}
then the line-element (\ref{3.9}) becomes:

\begin{equation}
\label{3.16.6}
d\bar s^2=h_2h_3dr^2+h_3h_1r^2d\theta^2+h_1h_2r2\sin^2\theta d\varphi^2 \quad
\end{equation}
with:

\begin{equation}
\label{3.16.7}
\partial_r h_1=-r^{-1}(2h_1-h_2-h_3),\ \ \partial_\theta h_2=-\cot \theta(h_2-h_3),
\ \ \partial_\varphi h_3=0
\end{equation}

Notice that the words Cartesian, Cylindrical and Spherical have a meaning when applied to the metrics $d\bar s^2$ because of the special relationship that we have established between this metric and $d\tilde s^2$, otherwise they would be meaningless.


\section{Time models}


Generally speaking, by a time model implementing an empiricist postulate, we mean any set of circumstantial conditions superimposed to a space model so that the coordinate time can
be identified with proper time along at least one of the world-lines of the congruence $\cal C$.

We discuss below two particular time models:

{\it Universal time}\,\footnote{Universal, Local and Ephemerides time names as used in this section are not meant to be exactly what they mean when commonly used in astronomy} .- Let us consider a particular hyper-surface $S_0$ transversal to $\cal C$  and consider the function:

\begin{equation}
\label{Time 1}
t^\prime(t,x^i)=\int^t_{t_0}\, d\tau
\end{equation}
where the proper-time integral is calculated along each world-line with space coordinates $x^i$ and $t_0$ is the intersection of it with $S_0$. The family of hyper-surfaces $t^\prime=cte.$ defines a universal time in the sense that one has $A=1$ everywhere, meaning that the time coordinate coincides with the proper-time on each of the world-lines of $\cal C$.

Since the hyper-surface is arbitrary Universal time and $f_i$ are only defined up to
the transformation:

\begin{equation}
\label{Time 2}
t^{\prime\prime}=t^\prime+\psi(x^j), \quad f_i^{\prime\prime}=f_i^\prime+\partial_i\psi(x^j)
\end{equation}
where $\psi$ is an arbitrary function of the space coordinates.

{\it Chorodesic Local time}.- A chorodesic of the line-element (\ref{0.1})  is a space-like integral of the differential equations (\cite{Bel1} and \cite{Bel2}):

\begin{equation}
\label{Time 3}
\frac{dp^0}{d\lambda}+\Gamma^0_{00}(p^0)^2
+2\Gamma^0_{0k}p^0p^k
+\Gamma^0_{ij}p^ip^j
=\frac{1}{2}A^{-1}\partial_t(A^{-2}\bar g_{ij})p^ip^j
\end{equation}

\begin{equation}
\label{Time 4}
\frac{dp^i}{d\lambda}+\Gamma^i_{00}(p^0)^2
+2\Gamma^i_{0k}p^0p^k
+\Gamma^i_{jk}p^jp^k=0
\end{equation}
the $\Gamma$'s being the Christoffel symbols and $p^\alpha=dx^\alpha/d\lambda$.

The important thing about the chorodesics of a congruence is that if one of them is orthogonal to one of de world-lines of the congruence then it is orthogonal to all of the world-lines that it crosses.

We define a Chorodesic Local time based on any particular world-line $W_0$ with space
coordinates $x^i_0$ of the congruence with the following intrinsic construction: let us consider the family $\cal F$ of hyper-surfaces generated by all the chorodesics issued from events on $W_0$ and being orthogonal to it that we consider to be the new hyper-surfaces $t^\prime=Const.$; they will be space-like in an open domain around $\cal C$. Choose an origin of time $t_0$ on $W_0$ and consider an event on a second
world-line $W$ with coordinates ($t,x^i$). Then the function $t^\prime$ will be:

\begin{equation}
\label{Time 5}
t^\prime(t,x^i)=\int^{t_1}_{t_0}\, d\tau,
\end{equation}
the proper-time integral being calculated along the world-line $W_0$ and $t_1(t,x^i)$ being the time coordinate of the intersection of the hyper-surface of $\cal F$ that contains the event ($t,x^i$) with $W_0$.
We shall have then $A^\prime\sim 1$, where the symbol $\sim$ means an equality on $W_0$.
And from $u_\alpha p^\alpha\sim 0$, $p^0\sim 0$ and $p^i$ being arbitrary on every event of $W_0$we get $f^\prime_i\sim 0$ and  $\partial_tf^\prime_i\sim 0$.

Let us consider the definition of the rotation rate of the congruence $\cal C$:

\begin{equation}
\label{Rot.1}
\Omega_{ij}=A(\partial_if_j+f_i\partial_tf_j-\partial_jf_i-f_j\partial_tf_i)
\end{equation}
If $\Omega_{ij}=0$ then the integrability conditions of the system of differential equations:

\begin{equation}
\label{Rot.1.1}
\frac{\partial t(t^\prime,x^j)}{\partial x^i}=f_i(t(t^\prime,x^j),x^k)
\end{equation}
are satisfied and therefore an adapted time transformation exists such that $f_i=0$ which means that
there exists a family $\cal I$ of hyper-surfaces orthogonal to $\cal C$. If a world-line
is intersected at by one hyper-surface of $\cal I$ at one event with
space coordinates $x_0^i$ it will contain all curves that are orthogonal at $\cal C$ and therefore it will contain also all the chorodesics issued from $x_0$ which means that the family of hyper-surfaces $\cal I$ coincide with the family of chorodesics $t^\prime=Const.$

Things are more complicated if $\Omega_{ij}\not=0$. We offer below the calculation of the
first and second derivatives of $f_i(t^\prime,x^j)$ on the world-line corresponding to
the chorodesic time $t^\prime$.

From $dp^0/d\lambda \sim 0$ we have:

\begin{equation}
\label{Time 5.0}
\partial_i f^\prime_j+\partial_j f^\prime_i\sim 0
\end{equation}
and also from $d^2p^0/d\lambda^2 \sim 0$:

\begin{equation}
\label{Time 5.0.1}
\partial_{ij} f^\prime_k+\partial_{jk} f^\prime_i+\partial_{ki} f^\prime_j\sim 0
\end{equation}
From the definition of $\Omega_{ij}$ above get:

\begin{eqnarray}
\label{Rot.2a}
\Omega_{ij}&\sim&\partial_i f_j-\partial_j f_i  \\
\label{Rot.2b}
\partial_k \Omega_{ij}&\sim&\partial_kA(\partial_i f_j-\partial_j f_i)+
\partial_{ki}f_j-\partial_{kj} f_i
\end{eqnarray}
or:

\begin{equation}
\label{Rot.3}
\partial_k \Omega_{ij}-\partial_kA\Omega_{ij}\sim \partial_{ki}f_j-\partial_{kj} f_i
\end{equation}
From (\ref{Time 5.0}) and (\ref{Rot.2a})we get:

\begin{equation}
\label{Rot.4}
\partial_if_j\sim\frac12 \Omega_{ij}
\end{equation}
and from (\ref{Time 5.0.1}) and (\ref{Rot.2b}) after symmetrization of the indices $j$ and $k$ we get:

\begin{equation}
\label{Rot.5}
\partial_{jk}f_i\sim -\frac13 (\partial_k \Omega_{ij}-\partial_kA\Omega_{ij}-
\partial_j \Omega_{ik}-\partial_jA\Omega_{ik})
\end{equation}

{\it Ephemerides time}.- Besides the particular case considered above where
the congruence $\cal C$ was integrable, a second case deserves particular attention; namely the case where the frame of reference is stationary. As it is well known, when the congruence $\cal C$ is a Killing congruence, then an Ephemerides time coordinate can be chosen such that:

\begin{equation}
\label{Time 6}
\partial_t A=0, \quad \partial_t f_i=0, \quad \partial_t \bar g_{ij}=0;
\end{equation}
and it is defined up to the transformation:

\begin{equation}
\label{Time 6.1}
t^{\prime\prime}=\lambda t^\prime+\psi(x^i)
\end{equation}
where $\lambda$ is a constant and $\psi$ is an arbitrary function of the space coordinates.

The ephemerides time model is compatible with the local chorodesic time model, \cite{Bel3}.
On the other hand the ephemerides time model is not in general compatible with universal time.


\section{Spherically symmetric models}

Let us consider the following spherically symmetric line-element:

\begin{equation}
\label{1.1}
ds^2=-A(t,r)^2(-dt+f(t,r)dr)^2 \\ [1ex]+A(t,r)^{-2}d\bar s^2
\end{equation}
where:

\begin{equation}
\label{1.2}
d\bar s^2=\left(1-k\frac{r^2}{4}\right)^{-2}(B(t,r)^2dr^2+B(t,r)C(t,r)r^2d\Omega^2)
\end{equation}
and:

\begin{equation}
\label{1.3}
d\tilde s^2=\left(1-k\frac{r^2}{4}\right)^{-2}(dr^2+r^2d\Omega^2)
\end{equation}
Equivalently this line-element is often written in several other familiar forms after a radial coordinate transformation $r=r(r^\prime)$:

\begin{equation}
\label{1.2.1}
d\bar s^2=\frac{{dr^\prime}^2}{1-k{r^\prime}^2}+{r^\prime}^2d\Omega^2
\end{equation}
or:

\begin{equation}
\label{1.2.2}
d\bar s^2={dr^\prime}^2+|k|\sinh(\sqrt{|k|}r^\prime)^2d\Omega^2)  \quad \hbox{if} \quad  k<0
\end{equation}
and:

\begin{equation}
\label{1.2.3}
d\bar s^2={dr^\prime}^2+|k|\sin(\sqrt{|k|}r^\prime)^2d\Omega^2)  \quad \hbox{if}\quad   k>0
\end{equation}

Since this metric has constant curvature $k$ the free mobility postulate has been built in the model.

To implement a model of space in the sense of the preceding section, Eq. (\ref{0.15}), the following equation has to be satisfied:

\begin{equation}
\label{1.8}
\frac{\partial C}{\partial r}+f\frac{\partial C}{\partial t}+C\frac{\partial f}{\partial t}=
\frac{2}{r}(B-C)\frac{4+kr^2}{4-kr^2}
\end{equation}
when the line-element (\ref{1.3}) is used and:

\begin{equation}
\label{1.8.2x}
\frac{\partial C}{\partial r}+f\frac{\partial C}{\partial t}+C\frac{\partial f}{\partial t}=
\frac{2}{r}(B-C)
\end{equation}
when one of the other three mentioned forms is used.

Let us consider a general adapted time-transformation:

\begin{equation}
\label{1.4}
t=t(t^\prime,r)
\end{equation}
Under such a time transformation we have:

\begin{equation}
\label{1.5}
A^\prime(t^\prime,r)=A\frac{\partial t}{\partial t^\prime}
\end{equation}
Similarly:

\begin{equation}
\label{1.6}
B^\prime(t^\prime,r)=B\frac{\partial t}{\partial t^\prime}, \quad  C^\prime(t^\prime,r)=
C\frac{\partial t}{\partial t^\prime}
\end{equation}
and also:

\begin{equation}
\label{1.7}
f^\prime(t^\prime,r)=\left(\frac{\partial t}{\partial t^\prime}\right)^{-1}\left(f-\frac{\partial t}{\partial r}\right)
\end{equation}

While the Universal time model and the Ephemerides time model when the frame of reference is stationary can be implemented as usual, the Chorodesic time model can be conveniently simplified requiring only the radial curves $t^\prime=Cte.$ to be chorodesics.  This leads immediately to the condition $f^\prime(t^\prime,r)=0$, that is known to follow also from more general arguments.


\section*{Acknowledgments}

I gratefully acknowledge the careful reading of this manuscript by J. Mart\'{\i}n
and the suggestions he made to me to improve it.



\section*{Appendix A}


These are two excerpts from \cite{Russell}.
\vspace{1cm}

{\it 72. (3)... This argument leads us to Land's distinction of physical and geometrical rigidity.
The distinction may be expressed and I think it is better expressed by distinguishing
between the conceptions of matter proper to Dynamics and to Geometry respectively. In
Dynamics, we are concerned with matter as subject to and as causing motion, as affected by
and as exerting force. We are therefore concerned with the changes of spatial configuration
to which material systems are liable: the description and explanation of these changes is
the proper subject-matter of all Dynamics. But in order that such a science may exist, it is obviously necessary that spatial configuration should be already measurable. If this were
not the case, motion, acceleration and force would remain perfectly indeterminate. Geometry,
therefore, must already exist before Dynamics becomes possible: to make Geometry dependent
for its possibility on the laws of motion or any of their consequences, is a gross
hysteron proteron.}
\vspace{1cm}

{\it ...Thus, to conclude: Geometry requires, if it is to be practically possible, some body or bodies which are either rigid (in the dynamical sense), or known to undergo some definite changes of shape according to some definite law. (These changes, we may suppose, are known by the laws of Physics, which have been experimentally established, and which throughout assume the truth of Geometry.) One or more such bodies are necessary to applied Geometry but only in the sense in which rulers and compasses are necessary. They are necessary as, in making the Ordnance Survey, an elaborate apparatus was necessary for measuring the base line on Salisbury Plain. But for the theory of Geometry, geometrical rigidity suffices, and geometrical rigidity means only that a shape, which is possible in one part of space, is possible in any other.}


\section*{Appendix B}


To state the Postulate of free mobility, when discussing the Fundamental principles
of geometry, it is usual to appeal to our immediate intuition of solids. We believe that
it is simpler, following the steps below, to appeal to our immediate intuition of flexible,
open threads, with two distinct end-points, that can not be elongated nor contracted unless submitted to an
``excessive" tension.

{\bf Definition 1}.- Two threads have the same length if they can be superimposed.

{\bf Definition 2}.- Let us select a thread $U$ to which we attribute the length $1$.

Then a thread has a fractional length $n/m$ if $m$ juxtaposed copies of it has the same
length as $n$ juxtaposed copies of $U$.

This is as much as it is necessary to speak about the quantity {\it Length}

{\bf Definition 3}.- A stretched thread is a thread submitted to an ``acceptable tension".

{\bf Axiom 1}.- Let $T$ be an stretched thread with length $L$ . Then no thread with length less
than $L$ can have the same origin and end as $T$.

{\bf Definition 4}.- A Simplex of dimension $d$ is a set of of $d+1$ distinct vertex joined by stretched threads, such that at each vertex coincide $d$ ends of $d$  different threads.

{\bf Axiom 2}.- In physical space the maximum value of $d$ is $3$ if we wish all the threads involved in the simplex to have the same length.

{\bf Free mobility postulate}.- Every simplex that exists at some location of physical space can be moved or duplicated anywhere else.

{\bf Definition 5}.- The distance from one point to another is the length of the shorter stretched thread
with one end at one point and the other end at the other point.

{{\bf Theorem} (Riemann, Helmholtz). If the geometry of physical space is Riemannian, i.e., if the distance between two points can be calculated by the integral along a geodesic of the
square-root of an elliptic quadratic form, then the Free mobility postulate requires the
curvature of space to be constant.


\begin{thebibliography}{9}

\bibitem{Russell} B.\ Russell, {\it An essay on the foundations of geometry}, Reprinted by Routledge London (1996); available at http://www.archive.org/details/117723764
\bibitem{Bel1} Ll.\ Bel in {\it Relativity in General}, Eds. J.\ Diaz, M.\ Lorente, Editions Fronti\`{e}res, 47 (1994); arXiv:gr-qc/1103.2509
\bibitem{Bel2} Ll.\ Bel and J.\ Llosa {\it Class. Quantum Grav.}, {\bf 12}, 1949 ( 1995)
\bibitem{Bel3} Ll.\ Bel in {\it Relativity and Gravitation in General}, Eds. J.\ Mart\'{\i}n, E.\ Ruiz, F.\ Atrio and A.\ Molina, World Scientific (1999); ArXiv:gr-qc/9812062

\end{thebibliography}
\end{document}